\documentclass[prl,aps,pacs,twocolumn]{revtex4}
\usepackage{epsfig}
\usepackage{bm}
\usepackage{color}

\newcommand{\B}[1]{{\bm{#1}}}
\newcommand{\C}[1]{{\mathcal{#1}}}
\usepackage[latin1]{inputenc}
\newcommand{\beq}{\begin{equation}}
\newcommand{\eeq}{\end{equation}}
\newcommand{\bea}{\begin{eqnarray}}
\newcommand{\eea}{\end{eqnarray}}
\begin{document}
\title{Finite-time Singularities in Surface-Diffusion Instabilities are Cured by Plasticity}
\author{Ting-Shek Lo, Anna Pomyalov, Itamar Procaccia and Jacques Zylberg}
\affiliation{Dept of Chemical Physics, The Weizmann Institute of Science, Rehovot 76100, Israel}
\begin{abstract}
A free material surface which supports surface diffusion becomes unstable when put under external non-hydrostatic stress. Since the chemical potential on a stressed surface is larger inside an indentation, small shape fluctuations develop because material preferentially diffuses out of indentations.  When the  bulk of the material is purely elastic one expects this instability to run into a finite-time cusp singularity. It is shown here that this singularity is cured by plastic effects in the material, turning the singular solution to a regular crack.
\end{abstract}
\maketitle

We address the problem of an amorphous material with a free surface on which the material
can diffuse such that the surface normal velocity is proportional to $\partial^2\mu/\partial b^2$ where
$\mu$ is the local chemical potential and $b$ is the parameterization of the interface. When the
material itself is purely elastic, this phenomenon leads to an instability which was termed
`thermal grooving' by Mullins \cite{Mullins} who discovered it. The phenomenon of grooving is
equally present in crystals, where it provides an annoying
mechanism for the failure of growing crystals \cite{BaliHeCr,GrinfElastStab}, as it does
in a range of amorphous solids that concern us here. Mullins considered the linear instability taking  into account only the curvature dependence of  the chemical potential. In fact, the chemical
potential along the interface is strongly dependent on the elastic energy;  linear stability analysis
taking both effects into account \cite{Srolovitz} reveals that the surface is stable for short
wavelengths but unstable for longer ones, with a usual 'fastest growing mode' whose wavelength
depends on the material parameters. The late stage of development of this instability was
initially studied by a handful of researchers, namely Asaro and Tiller \cite{AsaroTiller} and Grinfeld
\cite{Grinfeld} and followed by many \cite{Srolovitz, YangWeinan,
03BS,03BHP}. The result is that the instability runs into a finite-time singularity, with the
growing indentation forming a cusp. Clearly,  this often explored  \cite{Pimpinelli,GaoCycloid,YangWeinan} mathematical phenomenon cannot be
physical, and its discovery leaves open the question of the physical mechanism that
may cure the singularity. 

The question of what might cure the finite-time singularity in the Asaro-Tiller-Grinfeld (ATG) instability
remained dormant until recently Brener and Spatschek proposed that inertial effect in the velocity
of the moving boundary may tame the singularity \cite{03BS}. These authors pointed out that without
inertial effects the velocity of the tip $v$ appears in one dimensionless combintation, i.e. $vr_0^3/D$ where $r_0$ is the radius of the tip and $D$ the diffusion coefficient (of dimension length$^4$/time, and cf. Eq. (\ref{vn})). Therefore there is no mechanism to select $v$ or $r_0$, and as $r_0$ decreases without limit, $v$ increases without imit. Once inertial effects are taken into account the velocity appears also in the combination $v/v_R$ where $v_R$ is the Rayleigh wave speed. Thus a selection of both $v$ and $r_0$ can hapen. While clearly correct, the present authors stress that in many cases the surface diffusion is very slow, leading to small interface velocities which do not justify the incorporation of inertial terms. We focus here on such cases where the question of taming the cusp-singularities remains open.

In this Letter we propose that the generic mechanism for the taming of the ATG instability
may be plastic deformation in the stressed material, especially near the putative cusp. To test and demonstrate this proposition we will employ the recently proposed theory of elasto-plastic dynamics in amorphous systems \cite{07BLP}. To this theory, which is valid in the bulk of the material, we
couple the surface diffusion, allowing the chemical potential to take its stress dependence
from the elasto-plastic theory. For concreteness we choose to explore this interesting physics
on the inner surface of a circular hole which is stressed at infinity in a radial fashion.
The surface diffusion modifies the shape of
the slightly perturbed circular hole, leading eventually to a highly non-linear morphology. With a
sharpening interface due to the surface diffusion instability, stresses in the bulk
increase rapidly, exceeding at some point in time the yield stress of the material, triggering plastic flows which are dissipated by the exertion of plastic work  \cite{07BLP,Cavitation,STZHole}. It is interesting to observe the
coupling of both processes, namely surface diffusion and plasticity, as they become
competitive and of opposite influence on the morphology, to a point where the finite time
singularity is removed. In addition to shedding new light on the late stage of the ATG instability we find that the elasto-plastic theory employed here, which is sensitive to the
plastic properties of matter, allows a natural understanding of this a-priori seemingly hard problem.

The model system that we consider here is an infinite 2-dimensional isotropic elasto-plastic sheet with a
hole in the center whose radius is $R(\theta)$. For $r(\theta)<R(\theta)$ the system is void, whereas the elasto-plastic material occupies the region $r(\theta)\ge R(\theta)$ . The boundary is traction free, meaning that on the boundary $\sigma_{ij}n_j=0$ where $\B \sigma$ is the stress tensor and $\B n$ is the unit normal vector. 
The equations of acceleration and continuity are exact, reading:
\begin{eqnarray}
\rho\frac{\mathcal{D}\B v}{\mathcal{D}t}&=&\B \nabla\cdot\B \sigma  \label{acceleration}\\
\frac{\C D \rho}{\C D t}&=&-\rho \B \nabla\cdot\B v
\label{continuity}
\end{eqnarray}
Here the full material derivative $\mathcal D$ is defined for an arbitrary tensor $\bf{A}$ as:
\begin{equation}
\frac{\C D\B A}{\C D t}= \partial_t \B A +
\B v\cdot \B \nabla \B A+\B A\cdot \B \omega-\B \omega \cdot  \B A,
\end{equation}
where $\B \omega$ is the spin tensor $\omega_{ij}\equiv\frac{1}{2}\Big(\frac{\partial v_i}{\partial
x_j}-\frac{\partial v_j}{\partial x_i}\Big)$. Reading Eq. 
(\ref{acceleration}) in radial coordinates we get:
\begin{widetext}
\begin{eqnarray}
\rho\Big(\frac{\partial v_r}{\partial t}\!+\!v_r\frac{\partial
v_r}{\partial r}\!+\!\frac{v_\theta}{r}\frac{\partial v_r}{\partial
\theta}-\frac{v^2_\theta}{r}\Big)&\!\!=\!\!&\frac{1}{r}\frac{\partial \tau}{\partial
\theta}-\frac{1}{r^2}\frac{\partial}{\partial
r}(r^2s)-\frac{\partial p}{\partial
r}\ , \label{1}\\
\rho\Big(\frac{\partial
v_\theta}{\partial t}\!+\!v_r\frac{\partial v_\theta}{\partial
r}\!+\!\frac{v_\theta}{r}\frac{\partial v_\theta}{\partial
\theta}+\frac{v_rv_\theta}{r}\Big)&\!\!=\!\!&\frac{\partial \tau}{\partial
r}+\frac{1}{r}\frac{\partial s}{\partial
\theta}-\frac{1}{r}\frac{\partial p}{\partial
\theta}+\frac{2\tau}{r} \ . \label{2}
\end{eqnarray}
\end{widetext}
Here $s$ and $\tau$ are defined via the transformations
\begin{eqnarray}
\sigma_{rr} = s_{rr}-p\ ; \quad \sigma_{\theta\theta} &= &s_{\theta\theta}-p\ ; \quad \sigma_{r \theta}=\tau\ , \nonumber\\
\sigma_{rr} &=&-s_{\theta\theta}=-s \ .
\end{eqnarray}
Below our velocities are sufficiently small to allow neglecting the nonlinear terms $-v_\theta^2/r$
and $v_rv_\theta/r$. On the other hand nonlinear terms containing derivatives are retained,
since the derivatives are large.

The velocity at the interface, $\dot R(\theta)$, reads:
\begin{equation}
\frac{\partial R}{\partial t}= \frac{v_\theta}{R}
\partial_\theta R+v_r
\end{equation}

When the surface evolves, the stresses in the bulk evolve
accordingly. A fundamental assumption of our elasto-plastic theory is that the
total rate of deformation $\B D^{tot}\equiv \case{1}{2}[\B \nabla \B v+(\B \nabla \B v)^\dag]$ can be a represented as a linear combination of its elastic and
plastic components \cite{STZHole}:
\begin{equation}
\B D^{tot} \equiv \B D^{el}+\B D^{pl}
\end{equation}
Here the elastic contribution, $D^{el}$, is assumed to be linearly dependent on the stress (linear elasticity)

\begin{equation}
D^{el}_{ij}\equiv\frac{\mathcal{D}\epsilon_{ij}}{\mathcal{D}t} \quad
\epsilon_{ij}=-\frac{p\delta_{ij}}{2K}+\frac{s_{ij}}{2\mu}
\end{equation}
where $K$ and $\mu$ are the 2-dimensional bulk and shear moduli
and $p$ and $s_{ij}$ are the pressure and the
deviatoric stress tensor, respectively. The plastic rate of deformation, $D^{pl}$, is determined by a set of
internal fields which are discussed at length in \cite{07BLP} where the elasto-plastic theory is presented in 
detail. For the purpose of this Letter it is enough to state that the tensorial field $\B m$ acts as a 'back-stress'
due to plastic deformations, and the scalar field $\chi$ is the effective temperature that controls the amount
of configurational disorder in the elasto-plastic materials. The constitutive relations that were derived for
these fields read
\begin{eqnarray}
D^{pl}_{ij}&=&e^{-\frac{1}{\chi}}\mathcal{C}(\tilde{s})\Big(\frac{s_{ij}}{\tilde{s}}-m_{ij}\Big)
\\
\mathcal{D}m_{ij}/\mathcal{D}t&=&2e^{\frac{1}{\chi}}D_{ij}^{pl}-
\Gamma(s_{ij}, m_{ij})m_{ij}
\\ \nonumber \\
\mathcal{D}\chi/\mathcal{D}t&=&e^{-\frac{1}{\chi}}\Gamma(s_{ij},
m_{ij})(\chi_\infty-\chi)\\
\Gamma(s_{ij},m_{ij})&=&s_{ij}D^{pl}_{ij}/e^{-\frac{1}{\chi}}
\\
\mathcal{C}(\tilde{s})&=&\frac{e^{-\tilde s}(2+\tilde s)+\tilde s -2}{1+e^{-6(\tilde s-1.5)}}
\end{eqnarray}
In these equations all the stresses were normalized by the yield stress of the material $s_y$, using
$\tilde{s}=\sqrt{s_{ij}s_{ij}/2s^2_y}$. The function $\C C(\tilde s)$ has been chosen to make
the material relatively brittle.

To this theory we need to couple now the surface diffusion, expressed in terms of the 
normal velocity, $v_n(\theta)$ on the boundary. Without the effects of elasto-plasticity in the bulk
the normal velocity satisfies 
\begin{equation}
v_n=-\frac{D_s\Omega^2\delta}{k_BT}\frac{\partial^2 \mu}{\partial
b^2}  \label{vn}
\end{equation}
where $D_s$,
the surface diffusion constant, $\Omega$, particle volume and
$\delta$, the number of particles per unit area. In solving the coupled problem the total normal velocity should be computed as a {\em sum} of this contribution and the one coming from Eqs. (\ref{1}) and (\ref{2}).

The chemical potential on the boundary, $\mu(\theta)$, is associated
on the one hand with the destabilizing curvature and
on the other hand with the stabilizing surface energy
\cite{GrinfElastStab}:
\begin{equation}
\mu=\mu_0-\gamma\kappa+{\cal E} \ . \label{chempot}\\
\end{equation}
Here $\mu_0$ is the chemical potential of the unperturbed surface,
the curvature $\kappa(\theta)= (R^2+2R'^2-RR'')/(R^2+R'^2)^{3/2}$
and the strain energy density ${\cal
E}=\frac{1}{2}\sigma_{ij}\epsilon_{ij}$ \cite{LL}. $\gamma$ is the
surface energy.

In terms of these effects on the chemical potential one derives the equation for the
normal velocity due to surface diffusion alone:
\begin{equation}
v_n=-\frac{D_s\Omega^2\delta}{k_BT} \frac{\partial^2 \mu}{\partial
b^2} \Big[\Big(-\gamma\kappa(\theta)\Big)+ \Big(\frac{(1-\nu^2)}{2E}
\sigma_{ij}^2(\theta)\Big)\Big] \ ,
\end{equation}
where $\nu$ is the Poisson ratio and $E$ Young's Modulus. After non-dimensionalization
and projecting from $v_n$ to $v_r$ one ends up with the following equation in terms
of dimensionless quantities (denoted with tilde):
\begin{equation}
\tilde v_r= -\partial_{\theta} \frac{1}{\sqrt{\tilde R^2+\tilde R'^2}}
\partial_{\theta} \Big[-\tilde\kappa(\theta)+
\frac{(1-\nu^2)}{2} {\tilde\sigma}_{ij}^2
(\theta)\Big] \label{Vy}
\end{equation}

As noted,  we need to couple Eqs. (\ref{1} ),(\ref{2})  and (\ref{Vy}) to be solved together, such that the surface normal velocity is made from the sum of contributions coming from the bulk dynamics and the surface dynamics respectively. This should be done while keeping the traction free
boundary conditions and the initial condition of a pure elastic solution of a slightly perturbed circle.
In practice we used the fact that the elastic response is the fastest process in this problem. 
Accordingly we solved at each iteration first the elastic part of the model (with $D^{pl}=0$) to find the stress fields which is in agreement with the given interface, without taking into account any plastic deformation. Second, elasto-plastic relaxation was allowed to take place, until the 
system reached elasto-plastic equilibrium, allowing the interface to change. Here 'equilibrium' means
that $D^{pl}$ is smaller than $10^{-4}$. Lastly, a step of surface diffusion was
allowed to take place using an adaptive time step such as to bound the maximal movement of the boundary by $10^{-4}$. The last step changes the morphology of the boundary again, necessitating a re-calculation of the elastic fields around the new boundary, etc. Since the effect on the velocity of the 
interface in the last two steps  is additive, these steps (being infintesimal) could be also done simultaneously with impunity. We chose to separate the last two steps since the plastic and surface diffusion processes are non-dimensionalized independantly and have
different normalization values of characteristic times and stresses.

In order to realize an infinite sheet it is convenient to transform 
the $(r,\theta)$ coordinate system through
a conformal transformation to a new, finite domain $(\zeta, \theta)$ with $\zeta\in [0,1]$:
\begin{equation}
\zeta(\theta)=R(\theta,t)/r \ \ .
\end{equation}
In the finite space all the derivatives are computed using finite differences
and are redefined for the transformed space using the chain rule:
$\partial_{x_i}=\partial_{x_i}x_k\partial_{x_k}$.
Explicitly
\begin{equation}
\frac{\partial}{\partial r}\to-\frac{\zeta^2}{R}\frac{\partial}{\partial \zeta}\ , \quad
\frac{\partial}{\partial \theta}\to\frac{R'}{R}\zeta\frac{\partial}{\partial
\zeta}+\frac{\partial}{\partial{\theta}}\ , \quad
\frac{\partial}{\partial t}\to\frac{\dot
R}{R}\zeta\frac{\partial}{\partial \zeta}+\frac{\partial}{\partial
{t}}
\end{equation}
At $\zeta=0$ all the derivatives vanish and on the boundary $\zeta=1$
the time derivatives are estimated by linear extrapolation in the $\zeta$ direction. For the sake of
numerical stability small viscosity terms were added to the
acceleration components. The coupled equations were solved using $K=100$  and
$\mu=50$. Since the speed of sound is orders of magnitude larger than the velocity of the
interface, we could safely neglect the effects of Eq. (\ref{continuity}), giving up on seeing
the sound waves. The initial condition on the perturbed circle were
\begin{equation}
R(t=0,\theta) = 1+ 0.01 \cos 2\theta \ , \label{R0}
\end{equation}
the stress at infinity was chosen to be $\sigma_\infty = 0.9 s_y$. The Poisson ratio $\nu$ in Eq. (\ref{Vy})
is 1/3. The surface diffusion equations contain fourth order derivatives, calling
for spectral techniques for sufficiently stable evaluation. All the other derivatives were
computed by finite differences. 

\begin{figure}
\centering
\includegraphics[width=0.50\textwidth]{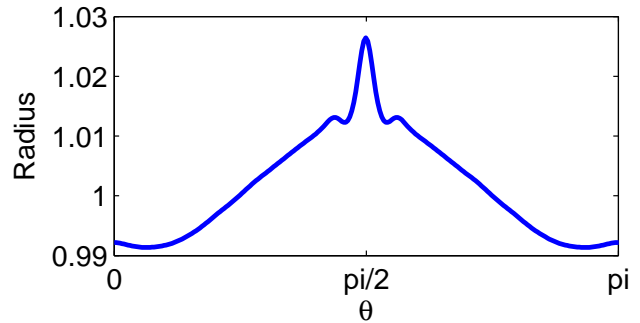}
\caption{A typical half profile of the stressed interface under the action of surface diffusion and plastiticity.
To the bare eye the effect of plasticity is not seen here, and one needs to compare elastic and plastic
solution in Figs. \ref{fts} and \ref{plastic}}
\label{interface}
\end{figure}

{\bf Results and Discussion}: The typical morphology of the unstable interface is shown in Fig. \ref{interface}
%%%%%%%%%%%%%%%%%%%%%%%%%%%%%%%%%%%%%%%%%%%%%%%%%%%
%%%%%%%%%%%%%%%%
%%%%%%%%%%%%%%%%%%%%%%%%%%%%%%%%%%%%%%%%%%%%%%%%%%%
\begin{figure}
\centering
\includegraphics[width=0.50\textwidth]{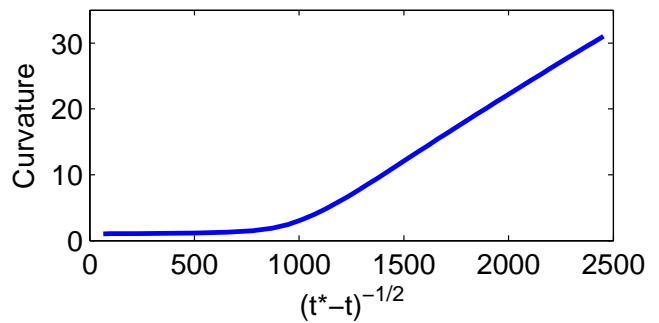}
\caption{The tip curvatures of the elastic solution. This solution appears to approach a singularity at $t^*= 5.37\times 10^{-4}$.}
\label{fts}
\end{figure}
%%%%%%%%%%%%%%%%
The elastic solution for the finite-time singularity is well
established and was reproduced in our numerics. The curvature $\kappa$ in the growing cusp first grows exponentially and rapidly switches to faster regime that agrees with the growth law
\begin{equation}
\kappa (t) \propto (t^*-t)^{-1/2} \ . \label{fast}
\end{equation}
To make this growth obvious we plotted the curvature of the elastic solution in Fig. \ref{fts}
as a function of $(t^*-t)^{-1/2}$ with $t^*= 5.37\times 10^{-4}$. Once plasticity is allowed to intervene, it
prevents the finite-time singularity by blunting the tip and by dissipating the stress. 
%%%%%%%%%%%%%%%%%%%%%%%%%%%%%%%%%%%%%%%%%%%%%%%%%%%
\begin{figure}
\centering
\includegraphics[width=0.50\textwidth]{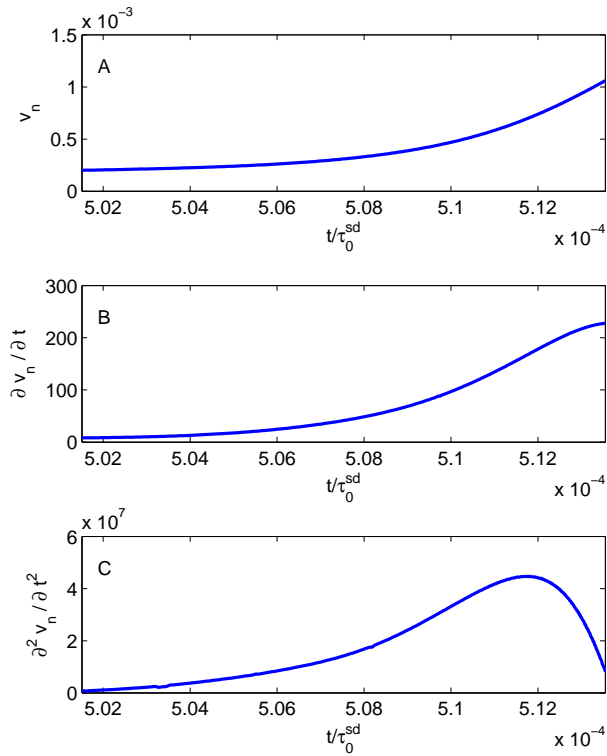}
\caption{The tip velocity (panel A), its first and second time derivatives (panels B and C respectively).
We see that the singularity is cured and the velocity decelerates due to the plastic effects.}
\label{plastic}
\end{figure}
%%%%%%%%%%%%%%%%

On the boundary, the smoothening of the interface in the vicinity of the
cusp via blunting is the ``cure" of the singularity. The
ever-increasing curvature occurring in the elastic solution is prevented in the plastic solution by the plastic flow induced by stress concentration. The avoidance of the singularity is
shown by the deceleration in the tip velocity, see Fig. \ref{plastic} panels B and C. 

It is important to stress that although the plasticity in the bulk succeeds to cure the finite-time
cusp-singularity, the role of surface diffusion is far from being negligible. Without it, the stressed
circle would remain stable to small shape fluctuations, as was demonstrated recently in \cite{08ELPS}.
The surface diffusion makes the circle unstable, and the instability results in the growth of a groove.
Without plasticity in the bulk the solution loses its meaning at $t=t^*$, whereas now, with plasticity 
playing its useful role, the solutions continue to exist at times $t>t^*$, in a form of a lengthening groove, or crack,
whose tip is protected from cusping by the plastic effects. At some point the crack will increase its
velocity due to the Griffith mechanism, and then the problem becomes inertial again.


\begin{thebibliography}{99}

\bibitem{Mullins} W. W. Mullins,
J. Appl. Phys.,{\bf 28}(3), 333-339 (1957).

\bibitem{BaliHeCr} R. H. Torii and S. Balibar,  J. Low Temp. Phys., 1992,
89, 391

\bibitem{GrinfElastStab} M. A. Grinfeld, J. Nonlinear Sc., 1992, 3, 35.

\bibitem{Srolovitz} W. H. Yang, D. J. Srolovitz, Phys. Rev. Lett., 1993, 71(10),
1593.

\bibitem{AsaroTiller} R. J. Asaro and W. A. Tiller, Metall. Trans., 1972, 3, 1789.

\bibitem{Grinfeld} M. A. Grinfeld, 
Sov. Phys. Dokl., 1986, 31, 831.

\bibitem{YangWeinan} X. Yang, E. Weinan, 2002, 91(11), 9414.

\bibitem{03BS} E. A. Brener and R. Spatschek, Phys. Rev. E, 2003, 67, 016112.

\bibitem{03BHP}
F. Barra, M. Herrera and I. Procaccia, Europhys. Lett., {\bf 63}, 708 (2003).

\bibitem{Pimpinelli} A. Pimpinelli and J. Villain, {\em Physics of Crystal Growth},
(Cambridge University Press, New York, 1998).

\bibitem{GaoCycloid} C. H. Chiu and H. Gao,  Int. J. Solids Struct, 1993, 30, 2983.


\bibitem{07BLP} 
E. Bouchbinder, J. S. Langer and I. Procaccia, Phys. Rev. E.,
{\bf  75}, 036107, 036108 (2007)

\bibitem{Cavitation} E. Bouchbinder, T. S. Lo and I. Procaccia, "Dynamic
Failure in Amorphous Solids via a Cavitation Instability", Phys. Rev. E, Rapid Communication, in press 

\bibitem{STZHole} E. Bouchbinder, J. S. Langer, T.-S. Lo and
I. Procaccia, Phys. Rev. E {\bf 76}, 026115 (2207).

\bibitem{LL} L.D. Landau and E.M. Lifshitz, {\em Theory of Elasticity},
3rd ed. (Pergamon, London, 1986).

\bibitem{08ELPS}
E. Bouchbinder, T.-S. Lo, I. Procaccia and E. Shtilerman, ``The Stability of an Expanding Circular Cavity and the Failure of Amorphous Solids", submitted to Phys. Rev. E.

\end{thebibliography}
\end{document}